\documentclass[12pt]{article}
\usepackage{epsfig}
\usepackage{graphicx}
\usepackage{times}
\usepackage{bm,amssymb}
\usepackage{subfig}
\usepackage{epsfig}
\usepackage{amssymb}
\def\beq{\begin{equation}}
\def\eeq#1{\label{#1}\end{equation}}
\def\eeqn{\end{equation}}

\def\beqa{\begin{eqnarray}}
\def\eeqa#1{\label{#1}\end{eqnarray}}
\def\eeqan{\end{eqnarray}}

\let\bar=\overbar

\def\Dslash{\not{\hbox{\kern-4pt $D$}}}
\def\dslash{\not{\hbox{\kern-2pt $\del$}}}

\def\msb{{\bar{\ssstyle M \kern -1pt S}}}

\usepackage{fancyhdr,graphicx}
\def\Title#1{\begin{center} {\Large {\bf #1} } \end{center}}

\begin{document}

\Title{Growth of the Magnetic Field in Young Neutron Stars}
% and Supernovae}

\bigskip\bigskip

%+\addcontentsline{toc}{chapter}{{\it D. Blaschke}}
%+\label{BlaschkeDavid}

\begin{raggedright}

{\it 
Rodrigo Negreiros$^{1,2}$~~C. G. Bernal$^{1}$\\
%\thanks{\tt Email: blaschke@ift.uni.wroc.pl}
\bigskip
$^{1}$Instituto de Fisica, Universidade Federal Fluminense, Av. Gal. Milton Tavares de Souza s/n, Gragoata, Niteroi, 24210-346, Brazil.\\
\bigskip

}

\end{raggedright}

\begin{figure}
\centering
\includegraphics[scale=0.3]{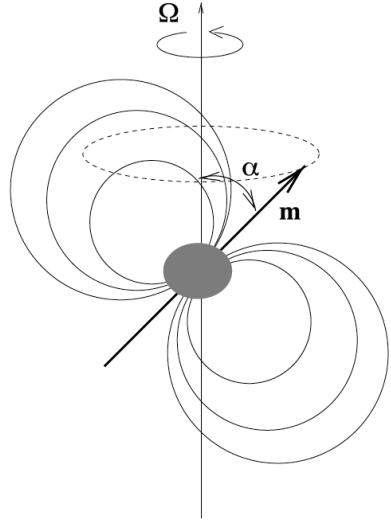}
\caption{Dipolar oblique rotator schematic: The pulsar rotate with angular velocity $\Omega$ and it has a magnetic moment $\mathbf{m}$ inclinated an angle $\alpha$ respect to the rotation axis.}
\label{Figure1}
\end{figure}

\label{section1}
At the end of their lives massive stars can form neutron stars in their womb, when they explode as core-collapse supernovae.
Neutron stars were proposed theoretically by Baade and Zwicky in 1934, but the observational confirmation of their existence only came in 1967 when they were discovered by Bell and Hewish as pulsars. The interpretation of these pulsars as rapidly rotating magnetized neutron stars was established by \cite{Gold1968}. This type of pulsars, in which the rotation of the neutron star is responsible for the observed luminosity, are known as Rotation-Powered Pulsars (RPP). 
The spin evolution of RPP ( in the canonical model) is well known and used to estimate ages and surface magnetic fields of neutron stars (see Fig. \ref{Figure1}). Here, we briefly summarize the main properties of the canonical model, in order to do a comparative analysis with our results.

The energy loss by radiation allows the neutron star to undergo a systematic spin-down (\cite{Manchester-Taylor1977}). If a pulsar spins down from an initial spin period $P_{0}$, then, the deceleration of the pulsar, $\dot \Omega$, is given by an empirical formula obtained by balancing the spin-down luminosity with the energy loss by radiation (a dipole magnetic field),

\begin{equation}
\dot{\Omega}=-k\Omega^{n},\quad k=\frac{2m^{2}\sin^{2}\alpha}{3Ic^{3}}, \label{sdlaw}
\end{equation}

\begin{table*}[h!]
\begin{scriptsize}
\centering

\begin{tabular}{cccccccc} 

\hline

Pulsar & Supernova & Period & Period derivative & Characteristic age & Current age & Braking index & References\\ 
name & remnant & $P$(s) & $\dot{P}$(s s$^{-1}$) & $\tau$(yr) & $t$(yr) & $n_{obs}$\\

\hline
\hline

B0531+21		& Crab		& 0.0331	& $4.23\times10^{-13}$	& 1240	& 960					& 2.51(1)		& Lyne 1988\\
J0537−6910	& N157B 		& 0.0161	& $5.18\times10^{-14}$	& 4930	& $2000_{-1000}^{+3000}$	& -1.5(1)	 	& Middleditch 2006\\
B0540-69		& 0540-69.3	& 0.0505	& $4.79\times10^{-13}$	& 1670	& $1000_{-240}^{+660}$		& 2.140(9)		& Nagase 1990\\
B0833−45		& Vela		& 0.0893	& $1.25\times10^{-13}$	& 11300	& $11000_{-5600}^{+5000}$	& 1.4(2)		& Lyne 1996\\
J1119−6127	& G292.2-0.5	& 0.408	& $4.02\times10^{-12}$	& 1610	& $7100_{-2900}^{+500}$		& 2.684(2)		& Weltevrede 2011\\
B1509−58		& G320.4-1.2	& 0.151	& $1.54\times10^{-12}$	& 1550	& $<21000$				& 2.839(3)		& Kaspi 1994\\
J1846−0258	& Kesteven 75	& 0.325	& $7.08\times10^{-12}$	& 729	& $1000_{-100}^{+3300}$		& 2.65(1)		& Livingstone 2007\\
J1734−3333	& G354.8-0.8	& 1.17	& $2.28\times10^{-12}$	& 8120	& $>1300$				& 0.9(2)		& Espinoza 2011\\
 
 \hline
 \end{tabular}
 \caption{Observational datas of young pulsars ($t\lesssim10\:\mathrm {kyr}$). The periods and period derivatives are taken from Manchester 2005. The braking indices obtained from observations, $n_{obs}$, are shown with their respective uncertainty in the last digit.}
 \label{Table1}
\end{scriptsize} 
 \end{table*}

\noindent where $k$ is a constant that accounts for structural factors of the pulsar, $n$ is the braking index and $\Omega=2\pi/P$ is angular velocity of the pulsar. The case $n=3$ correspond to purely dipolar radiation. Different values of $n$ would correspond to different processes of rotational energy loss. The characteristic age of the pulsar is given by,

{\small
\begin{equation}
\tau=\frac{\Omega}{(n-1)\dot{\Omega}}\left[\left(\frac{\Omega}{\Omega_{0}}\right)^{n-1}-1\right]=\frac{P}{(n-1)\dot{P}}\left[1-\left(\frac{P_{0}}{P}\right)^{n-1}\right]
\end{equation}
}

\noindent which is valid only when  $n\neq1$. The parameter $\Omega_{0}$ is the initial angular velocity at $t=0$. In the limit when $\Omega_{0}\gg\Omega$ (or equivalently $P_{0}\ll P$), the standard characteristic age of the pulsar is obtained: $\tau=-\Omega/(n-1)\dot{\Omega}=P/(n-1)\dot{P}=P/2\dot{P}$, assuming $n=3$. Traditionally, this expression is taken as the definition of characteristic age, despite the fact that $n$ can be different from 3, and is, in fact, so for all cases in which a stable, accurate value has been determined. In addition, if $n$ is a constant of the pulsar then the spin-down luminosity and the spin period evolve with time according to,

\begin{equation}
\dot{E}=\dot{E}_{0}\left(1+\frac{t}{\tau_{0}}\right)^{-\frac{(n+1)}{(n-1)}},\quad P=P_{0}\left(1+\frac{t}{\tau_{0}}\right)^{\frac{1}{n-1}}
\end{equation}

\noindent where $\tau_{0}=P_{0}/(n-1)\dot{P}_{0}$ is the initial spin-down timescale of the pulsar and $\dot{E}_{0}$ is the initial spin-down luminosity, which has roughly constant energy output,  $\dot{E}\thickapprox \dot{E}_{0}$, until a time $\tau_{0}$ beyond which evolve as $\dot{E}\propto t^{-(n+1)/(n-1)}$. Similary, $P\thickapprox P_{0}$ for $t\ll\tau_{0}$ and evolves at later times as $P\propto t^{1/(n-1)}$ (\cite {Gaensler-Slane2006}). Notice that changes in the moment of inertia are neglected as of now (we will discuss such possibilities in the second part of this paper).

The braking index $n$ can be obtained directly from timing observations, and its measurements are crucial to understanding the physics behind the pulsar. Its definition is obtained from the observables as, $n=\Omega\ddot{\Omega}/\dot{\Omega}^{2}=2-P\ddot{P}/\dot{P}^{2}$. 
The standard requirements for accurate measurements of $n$ are: (a) that the pulsar is slowing fast enough to measure $\ddot{\Omega}$, (b) known position of the pulsar ($\sim1$'') and (c) that the braking is not affected by other agents such as glitches, timing noise or low frequencies. The glitches are important for $t\sim 10-15 \:\mathrm {kyr}$. The timing noise varies from pulsar to pulsar,  but it is correlated with the spin-down and can affect $n$ if the observational data set is not very large (\cite{Manchester2006}). The difficulty of extending this analysis to older pulsars lies in the fact that for these pulsars unfortunately measurements of $\ddot{\Omega}$ require tens of years and the expected $\ddot{\Omega}$ are very small. Furthermore for older pulsars, the aforementioned effects, such as glitches and timing noise are more recurrent which make such measurements more challenging (\cite{Lyne-Graham-Smith2012}). Thus young pulsars ($t\lesssim10 \:\mathrm {kyr}$) are the best choice for measuring $n$. To date there have been only a few measurements for the braking index, and only for very few young pulsars. In all cases $n_{obs}<3$ (see the Table \ref{Table1} and references therein). 
For these young pulsar glitches are minor and the pulsar spin-down is faster. These results suggest that there should be present more complicated  processes in the pulsar decreasing $n$ from the expected value for the dipolar oblique rotator ($n=3$). The standard factors that may affect the braking index are (\cite{Muslimov-Page1996}): (a) multipolar electromagnetic radiation, $n\geq5$, (b) quadrupole gravitational radiation, $n=5$, (c) decay of the magnetic field, $n>3$, (d) radial deformation of the magnetic field lines, $1\leq n\leq3$, (e) relativistic winds, $n<3$, (f) transverse velocity of the pulsar, $n<3$. 

\noindent More exotic factors are: (g) Intense emission of neutrinos in the early evolution of the pulsar, $n<0$ (\cite{Alpar-Ogelman1990}), (h) Crustal movement of the neutron star by tectonic plates can produce $n>3$ or $n<3$ (\cite{Ruderman1991}), (i) Growth of the magnetic field due to  thermomagnetic instabilities in the crust of the neutron star, $n<3$ (\cite{Blandford-Romani1988}), (j) Growth of an intense magnetic field submerged on neutron star crust in the hypercritical accretion phase, which re-emerge by ohmic diffusion, $n<3$ (\cite{Muslimov-Page1996}, \cite{Bernal2010}, \cite{Bernal2013}, \cite{Vigano-Pons2012}), (k) Changes in the moment of inertia of the neutron star, $n<3$ (\cite{Weber1999}, \cite{Glendenning2003}, \cite{ho2012}).

\noindent The effects of alignment $\left(d\alpha/dt>0\right)$ or misalignment $\left(d\alpha/dt<0\right)$ on $n$ were calculated theoretically by \cite{Ghosh1984}. For a dipolar field, $n$ changes from its above canonical value of 3 by an amount $2\tau\left(d\ln\sin^{2}\alpha/dt\right)$ due to alignment or misalignment between the rotational and magnetic axes, where $\tau$ is the characteristic age introduced above. Nevertheless, observations of alignment or misalignment are very complicated to carry out because large amounts of data are required to make a statistical estimate. In addition, unfortunately, an exact expression for the electromagnetic (Poynting) flux energy loss as a function of the magnetic inclination angle $\alpha$ remains still elusive. For these reasons, in this work we do not take into account such effect (considering then $\sin\alpha\simeq1$) and we interpret the magnetic field $B$ as the component perpendicular to the stellar surface

In the present work we are interested in exploring the effects due to the growth of the magnetic field. The growth of the magnetic field in young pulsar are not a new idea, but the impact on the braking index and the early dynamic of the pulsar have not been fully explored.

\section{Growing magnetic fields in young pulsars}
\label{section2}
A newborn neutron star may be exposed to a hyperaccretion phase few moments after the supernova explosion that originated it. The paradigm is as follows: when the core-collapse supernova event take place the shock is still pushing its way through the outer layers of the progenitor, and if it encounters a density discontinuity, a reverse shock may be generated. Depending on its strength and on how far out it was generated, this reverse shock can induce strong accretion onto the newborn neutron star on a timescales of hours. Hypercritical accretion results ($\dot{m}>\dot{m}_{Edd}$, where $\dot{m}_{Edd}$ is the Eddington accretion rate), in which the photons are trapped within the accretion flow and the energy liberated by the accretion is lost through neutrino emission close to the neutron star surface. After the reverse shock hits the neutron star surface and rebounds, a third shock develops and starts moving outward against the infalling matter. Once this accretion shock stabilizes it will separate the infalling matter from an extended envelope in quasi-hydrostatic equilibrium.
\cite{Chevalier1989} argued in favor of such scenario of late accretion onto newborn neutron stars inside supernovae and developed an analytical model for the hypercritical regime. In such model, the neutrino cooling plays an important role in the formation of a quasi-hydrostatic envelope around the compact remnant.
\cite{Muslimov-Page1995} highlighted the physical conditions presents in the formation of neutron stars inside supernovae: convective envelope, hyperaccretion of material and submergence of the magnetic field on the stellar crust. With these suggestions, \cite{Geppert1999} presented simple 1D ideal MHD simulations of the effect of this post-supernova hypercritical accretion on the newborn neutron star to show that such magnetic field submergence could occur. The result was a rapid burial of the magnetic field into the neutron star crust.
Notice that an accreted mass of $\sim 0.001\:\mathrm{M_{\odot}}$ is enough to submerge the magnetic field (which corresponds to an accretion rate of $\sim100\:\mathrm{M_{\odot}}\:\mathrm{yr^{-1}}$ in few hours).
It is widely accepted that the origin of neutron star magnetic fields is still an unsolved problem (\cite{Spruit2008}, \cite{Spruit2009}). Until now, two main mechanisms are still competing: (a) a fossil field from the progenitor compressed during the core collapse (magnetic flux conservation and diamagnetism of MHD turbulence), and (b) a proto-neutron star dynamo (generated by precollapse or the short-lived postcollapse). Both models are used to explain the large variety of observed field strengths. However, currently is accepting the idea of magnetic field submergence by hyperacretion onto newborn neutron stars and subsequent reemergence, because it explains better why a neutron star at birth has a very low surface magnetization.
In these scenarios, the magnetic field generation and/or adjustment process terminates few seconds after the neutron star’s birth (\cite{Muslimov-Page1995}).
Recently, \cite{Bernal2010}; \cite{Bernal2013}; performing 2D-3D MHD simulations with more refined detailed physical ingredients, showed that the magnetic field is submerged in the stellar surface regardless of their initial configuration or its strength.
Although many observed pulsars show clear evidence of strong magnetic fields (from radio pulsar with $10^{12}$ G to magnetars with $10^{15}$ G), lower magnetic fields are, however, found in millisecond pulsars (\cite{Phinney-Kulkarni1994}) and in neutron stars in low-mass X-ray binaries (\cite{Psaltis2006}). In such cases, hyperaccretion is thought to be the cause of the magnetic field reduction. In addition, there is a small group of neutron stars, found in young supernova remnants which exhibit little or no evidence for the presence of a magnetic field (\cite{Geppert1999}; \cite{Ho2011}).

After hyperaccretion stopped, the magnetic field could diffuse back to the surface and result in a delayed switch-on of a pulsar (\cite{Michel1994}; \cite{Muslimov-Page1995}). Depending on the amount of accreted matter, the submergence could be so deep that the neutron star may appear and remain unmagnetized for several centuries or millennia (\cite{Geppert1999}). This scenario was recently revisited by \cite{Ho2011} and \cite{Vigano-Pons2012} and applied to study the field evolution of the CCOs (Central Compact Object; \cite{Pavlov2004}) defined as X-ray sources with thermal-like spectra observed close to the centers of SNRs without any counterparts in radio and gamma wavebands. With blackbody temperatures of about a few hundred eV and luminosities in the range $10^{33} - 10^{34}$ erg s$^{-1}$, they present no evidence of a pulsar magnetic field. Several of such sources are known, including RCW103, Cas A, Pup A, and Kes 79 (\cite{Kaspi2010}).
In the case of magnetic fiel submergence, the numerical simulations were done using an ideal MHD scheme. This is possible because of the short duration of initial transient and due to the violence of the hypercritical regime. In contradistinction, the back-diffusion of the field after hyperaccretion stops is due to the finite electrical conductivity of the neutron star crust matter. As a result, the back-diffusion becomes dependent on the thermal evolution of the star because of the temperature dependence of the electrical conductivity. \cite{Muslimov-Page1995} and \cite{Vigano-Pons2012} found that when the hypercritical phase is over, the magnetic field can re-emerge by a diffusion process. The difusion time in such case is $\tau_{B}\sim10^{2-3}\:\mathrm{yr}$. This is the timescale for grow a magnetic field from a low value  $B\sim10^{8-9}\:\mathrm{G}$ to a high value $B\sim10^{12-13}\:\mathrm{G}$, which depend of the early thermomagnetic history of the pulsar, the amount of accreted matter, the initial magnetic flux distribution and the electric resistivity in the crust. 
On the other hand, \cite{Magalhaes2012} admit that pulsars are rotating magnetic dipoles with growth of the magnetic field, but they did not propose any intrinsic mechanism for such growth. With such assumption and based on data from the seven pulsars with known braking indices they predict ranges for the braking indices of other pulsars.

Following these ideas, it is possible to study analytically, in a first approach, the growth of the magnetic field when it re-emerges from the neutron star crust, and to follow its consequences on the pulsar spin-down. 

\subsection{The modified canonical model}
The magnetic field re-diffusion from the new crust (formed in the hypercritical regime) to the stellar surface, under a variety of complex physical conditions, is a problem that generally requires numerical modeling (\cite{Geppert1999}; \cite{Vigano-Pons2012}). However, valuable insight can be gained from purely analytical solutions, which exhibit the main properties of the spin-down dynamics and may be applicable to various scenarios where the pulsar magnetic field is weak. We think that this study is academically important because it allows us to explore the behavior of certain parameters of the pulsar, in the early phase, under a different approach and without need for intensive numerical work.

If the magnetic field has a temporal dependence through a growth function $f(t)$, then it is possible to express the magnetic field as $B(t)=Bf(t)$, where $B$ is the maximum magnetic field achieved in the saturation regime. This temporal dependence affect the function of torque $k$, which is considered constant in the simple dipolar oblique rotator model. A similar approach was proposed initially by \cite{Blandford-Romani1988} who proposed to replace $k$ by some $f(t)$ and them attempt to constrain $f(t)$ using the observed value of $n$, but they proposed no growth function because his approach was focused in the thermomagnetic instabilities in the neutron star crust. In the present case we are interested in to study the magnetic field growth due a ohmic diffusion process in the stellar crust when the hypercritical phase is over. In such case, the structural factor of the pulsar is time-depending through the magnetic field as, $k(B(t))=k(t)$, modifying the general law for the pulsar spin-down,

\begin{equation}
\dot{\Omega}=-k(t)\Omega^{n},\quad k(t)=kf(t) \label{modlaw}
\end{equation}
\begin{equation}
\dot{\Omega}=-k(t)\Omega^{n},\quad k(t)=kf(t) \label{modlaw}
\end{equation}

\noindent where $f(t)$ is an analytical function that allows the growth of the magnetic field from an low initial value to a maximum value in the saturation regime, and $k$ is the standard constant of the pulsar. Changes in the moment of inertia are neglected in this approach (see next section in which such changes are investigated).
Performing a similar treatment as in the canonical model, we find the relationships for the overall properties of the young pulsar with growth of the magnetic field: characteristic age, braking index, spin-down luminosity and period.

The characteristic age is found integrating the modified general law (\ref{modlaw}) as,

\begin{equation}
\tau=\frac{1}{f(t)}\left[\tau_{0}+\intop_{0}^{t}f(t)dt\right].
\end{equation}

\noindent Notice that, depending on the function $f(t)$ proposed, the characteristic age changes substantially from its canonical value.

 The braking index $n$ is found by a straightforward differentiation of the modified general law (\ref{modlaw}) as,

\begin{equation}
 n=n_{*}+\frac{\dot{f}(t)}{f(t)}\frac{\Omega}{\dot{\Omega}}=n_{*}-\frac{\dot{f}(t)}{f(t)}\frac{P}{\dot{P}},
\end{equation}

\noindent where the theoretical braking index is $n_{*}=3$ and $\dot{f}(t)=df/dt$ is the temporal derivative of $f(t)$. Notice that to obtain $n<3$ is required magnetic field growth through $\dot{f}(t)>0$, because $\dot{\Omega}<0$ always (the pulsar is slowing). 

In addition, in the modified canonical model, the spin-down luminosity and the period are given by,

\begin{equation}
\dot{E}=\dot{E}_{0}f(t)\left[\frac{\tau}{\tau_{0}}f(t)\right]^{-\frac{(n+1)}{(n-1)}}, \quad P=P_{0}\left[\frac{\tau}{\tau_{0}}f(t)\right]^{\frac{1}{n-1}}.
\end{equation}

\noindent In this case, the temporal evolution of the spin-down luminosity and rotation period look more complex that in the canonical model, and depend exclusively of the growth function $f(t)$.

\subsection{The growth function $f(t)$}

Regardless of which analytical function $f(t)$ is proposed, one of its main requirements is that for $t>\tau_{B}$, the canonical properties for the dipolar oblique rotator model (characteristic age, braking index, spin-down luminosity and period) are recovered. That is, when the magnetic field reaches the saturation regime, the evolution of the pulsar follows the canonical model. Thus, it is required that $f(t)$ allows the growth of the magnetic field rapidly, in a diffusion timescale $\tau_{B}$, from a low value to a canonical value. As already mentioned, magnetic field strengths inferred from the observed pulsar population range between $B\sim10^{8}\:\mathrm{G}$ for recycled (or millisecond) pulsar, $B\sim10^{12}\:\mathrm{G}$ for normal pulsars up to $B\sim10^{15}\:\mathrm{G}$ for magnetars. Therefore, the growth function must satisfy that $f(t\rightarrow0)= \epsilon$ and $f(t>\tau_{B})=1$,  where $\epsilon=B_{0}/B\ll1$ is a parameter related with the magnetic field strength $B$, in the saturation regime, and $B_{0}$ is the initial low magnetic field. An exponential growth function is most appropriate in this case, which satisfies the aforementioned requirements,

\begin{equation}
f(t)=\epsilon+\left[1-\exp\left(-\frac{t}{\tau_{B}}\right)\right]. \label{grow}
\end{equation}

\begin{figure*}
\centering
\subfloat[$B(t)$ for several $\epsilon$, fixing  $\tau_{B}$]{
\includegraphics[height=5.0cm]{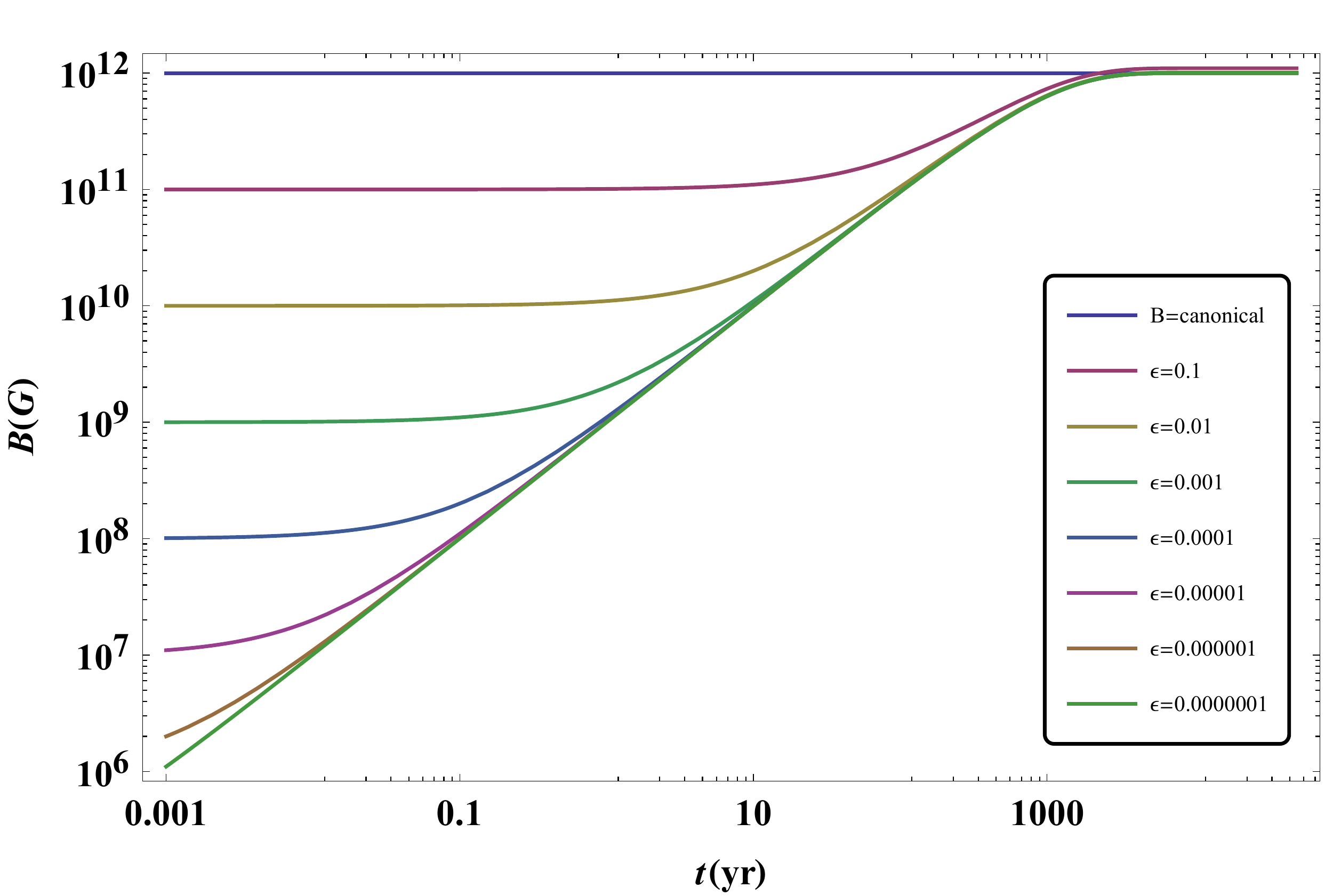}
\label{fig2a}
}
\quad %espaco separador
\subfloat[$\tau$ vs age for several $\tau_{B}$, fixing $\epsilon$]{
\includegraphics[height=5.0cm]{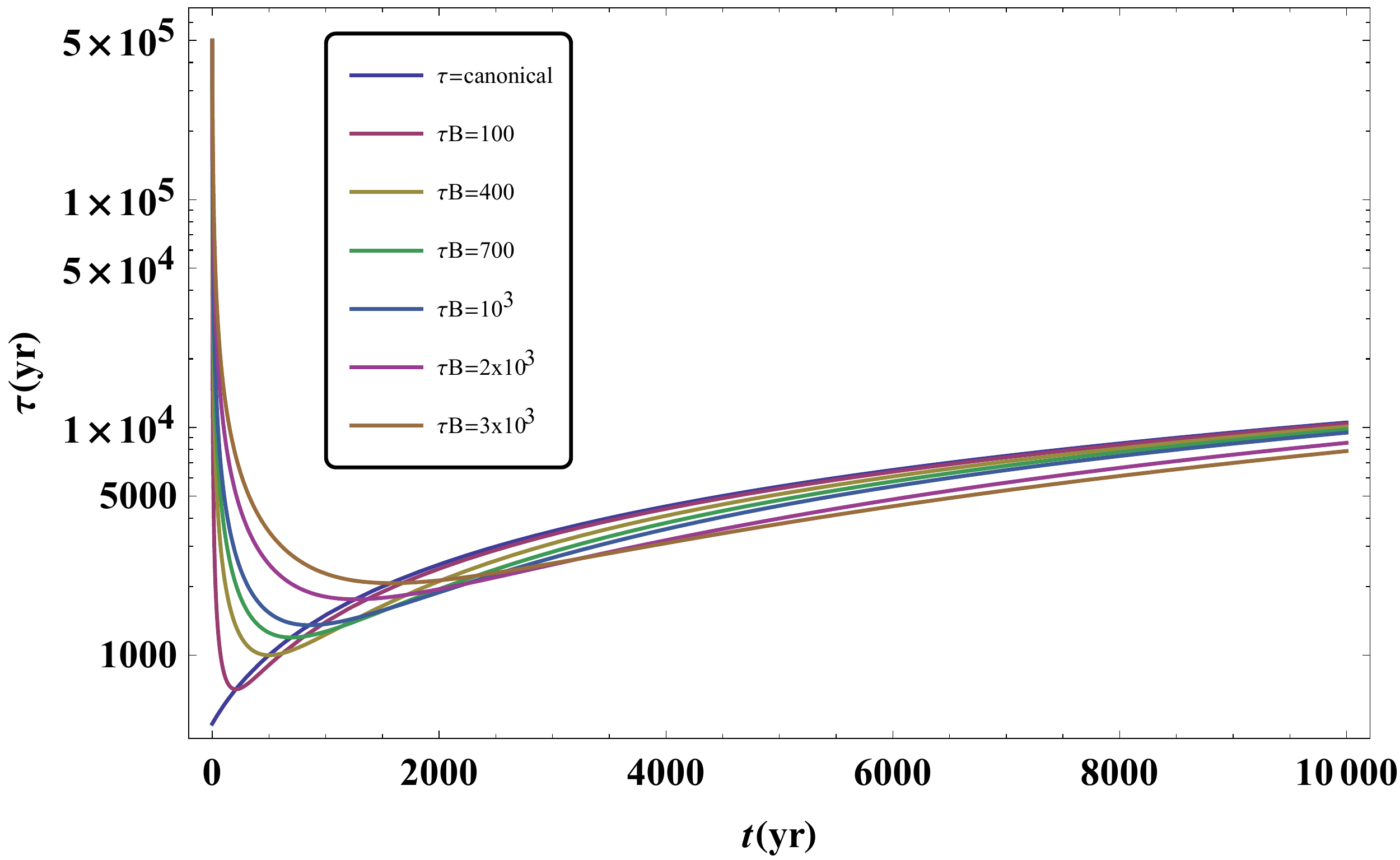}
\label{fig2b}
}
\caption{(color online) In (a) we show the growth of magnetic field through function $f(t)$, in logarithmic scale, varying $\epsilon$ for a fixed $\tau_{B}=10^{3}$ yr.  We show $B(t)$ for several values of $\epsilon$. The canonical model, $B=10^{12}$ G (or equivalently $f(t)=1$), is superimposed for comparison purposes. In (b) we show the modified characteristic age of the pulsar for a fixed value of $\epsilon=0.001$ and varying the diffusion timescale $\tau_{B}$. The canonical parameter $\tau$ is superimposed for comparison purposes.}
\label{Figure2}
\end{figure*}

\begin{figure*}
\centering
\subfloat[$\dot{E}$ for several $\tau_{B}$, fixing $\epsilon$]{
\includegraphics[height=5.0cm]{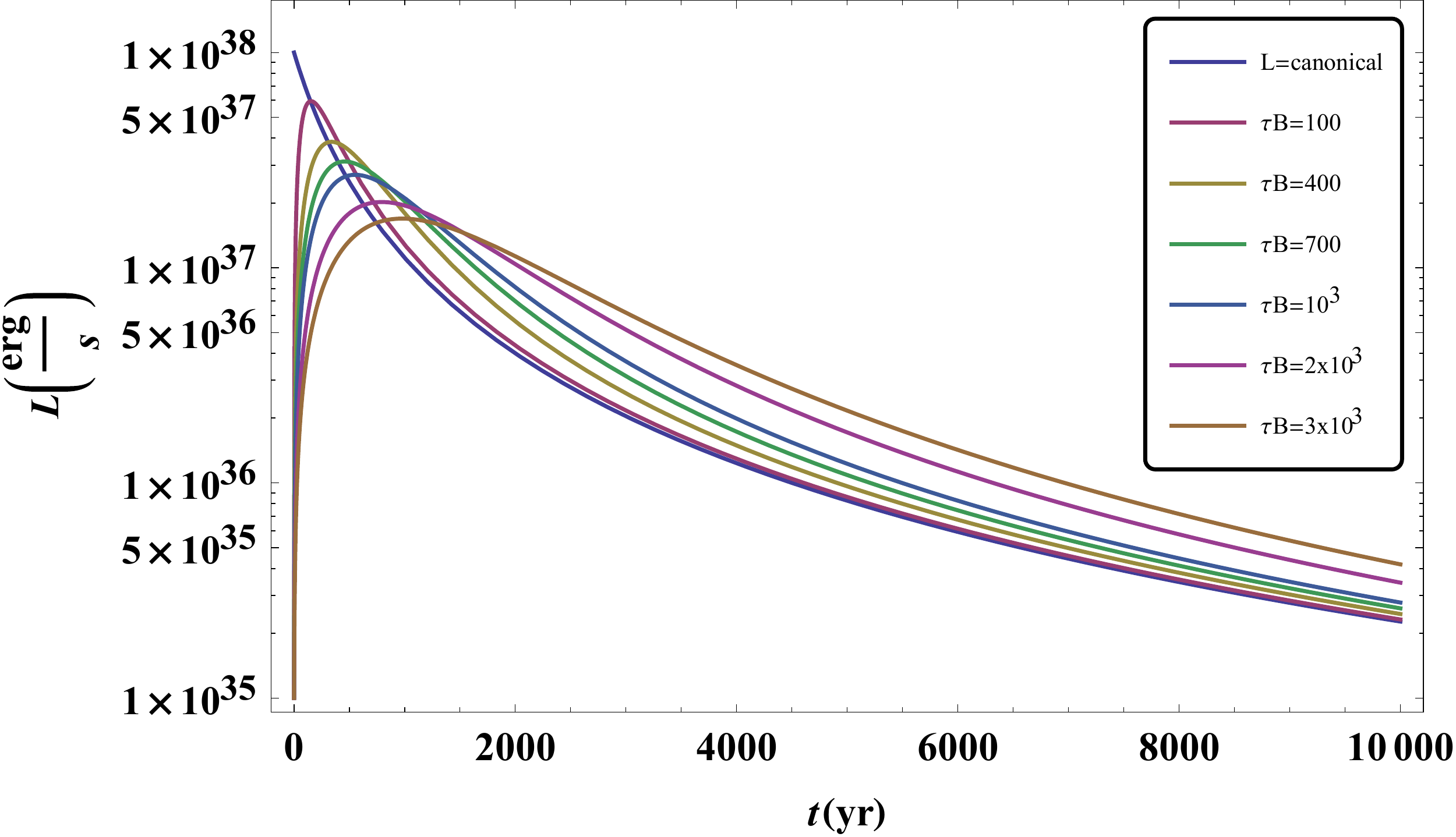}
\label{fig3a}
}
\quad %espaco separador
\subfloat[$n$ vs age for several $\tau_{B}$, fixing $\epsilon$]{
\includegraphics[height=5.0cm]{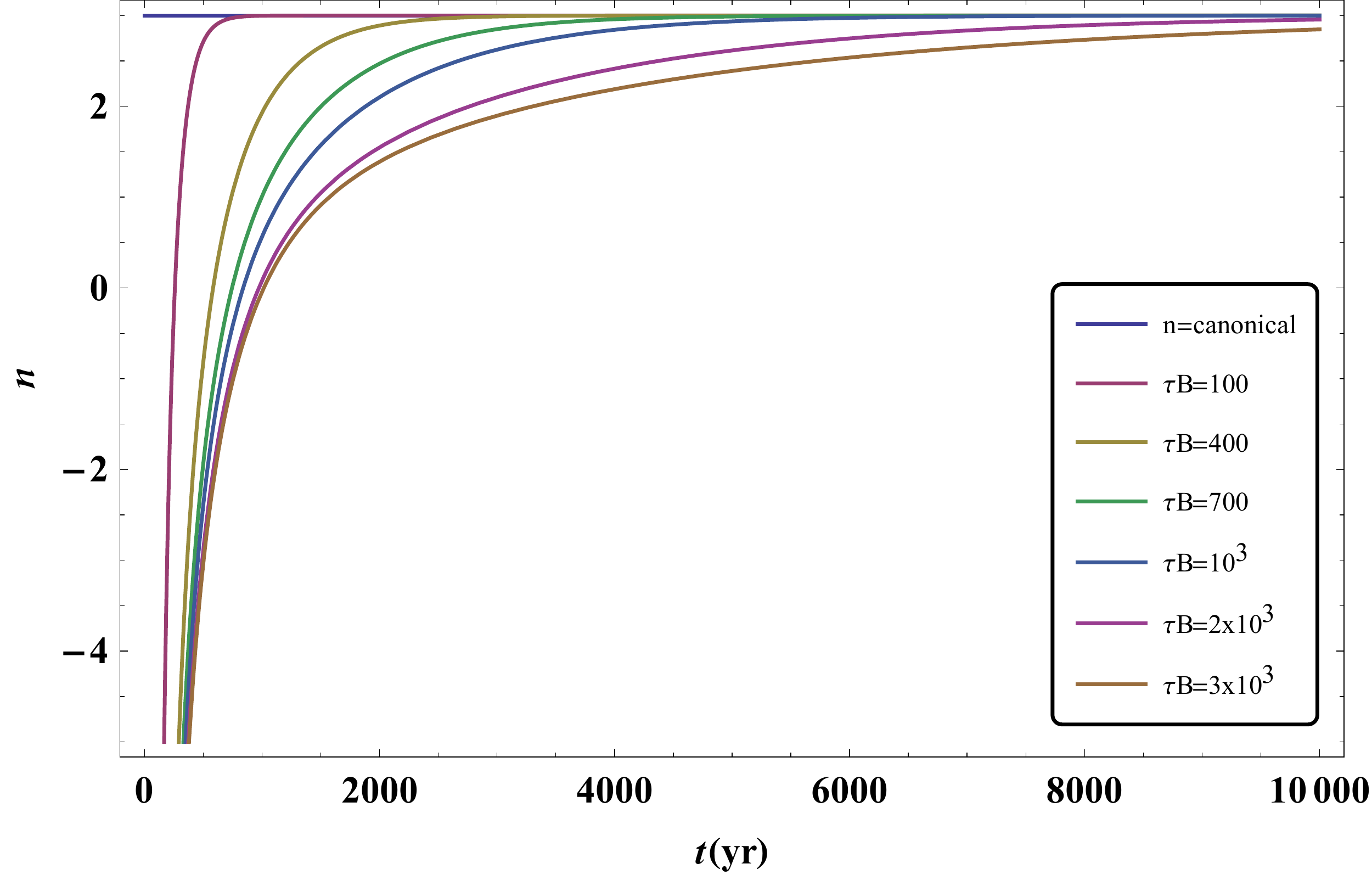}
\label{fig3b}
}
\caption{(color online) In (a) we show the modified spin-down luminosity of the pulsar for a fixed value of $\epsilon=0.001$ and varying the diffusion timescale $\tau_{B}$. In (b) we show the modified braking index of the pulsar for same parameters as (a). The canonical parameters, $\dot E$ and $n$, are superimposed for comparison purposes.}
\label{Figure3}
\end{figure*}

\noindent The parameters $\epsilon$ and $\tau_{B}$ involve the most relevant physical ingredients of the problem: magnetic fields, resistivity in the crust, MHD instabilities, submergence depths and more. However, the oversimplified growth function (\ref{grow}) match very well with those numerical magnetic curve obtained by \cite{Geppert1999} \& \cite{Vigano-Pons2012} for standard parameters. Fig.~\ref{Figure2} shows the evolution of the magnetic field through the growth function $f(t)$ and the characteristic age evolution for young pulsars. In (a) we show $B(t)$ for several values of $\epsilon$ and  for a fixed $\tau_{B}=10^{3}$ yr, to analyze its effect on the growth of the magnetic field. Notice the rapid magnetization of the pulsar. Due to the constriction on $\tau_{B}$, the magnetic field is saturated more rapidly for low values of $\epsilon$ and slower for higher values of the same parameter. That is, the magnetic field is almost constant during a time and may take longer to start the growth process for high values ​​of $\epsilon$, while for low values ​​of $\epsilon$ it may take less time to start such process of growth  (from $B=10^{5}\:\mathrm{G}$ to $B=10^{12}\:\mathrm{G}$). But in any case, the growth of the magnetic field is very efficient (occurring in a timescale $\tau_{B}$).
In (b) we show the characteristic age $\tau$  for several values of $\tau_{B}$ and for a fixed $\epsilon=0.001$. Notice that initially the modified $\tau$ is different from the canonical value, but after a diffusion timescale $\tau_{B}$ it evolve like the familiar canonical model. The main reason for this behavior is the parameter $\epsilon$ in the growth function $f(t)$ because when $t\rightarrow0,\:\tau\rightarrow\tau_{0}/\epsilon$, being $\tau_{0}$ the canonical initial spin-down timescale of the pulsar. That is, for an initial low magnetic field $B_{0}=10^{9}\:\mathrm{G}$, the corresponding initial characteristic age is three orders of magnitude higher that the canonical case, which is chosen as $\tau_{0}=500$ yr for this example, in particular.

In Fig. \ref{Figure3} we show both the spin-down luminosity and the braking index evolution. In (a) the modified spin-down luminosity is shown for several values of $\tau_{B}$ and for a fixed $\epsilon=0.001$. Notice that initially the modified spin-down luminosity $\dot E$ is different from the canonical value, but after a diffusion timescale $\tau_{B}$ it evolve like the familiar canonical model. Again, the reason for this behavior is the parameter $\epsilon$ in the growth function $f(t)$ because when $t\rightarrow0,\:\dot E\rightarrow\epsilon\dot E_{0}$, being $\dot E_{0}$ the canonical initial spin-down luminosity of the pulsar. That is, for an initial low magnetic field $B_{0}=10^{9}\:\mathrm{G}$ the corresponding initial spin-down luminosity is three orders of magnitude lower that the canonical case, which is chosen as $\dot E_{0}=10^{38}$ erg s$^{-1}$ for the present example. In (b) we show the modified braking index $n$ for several values of $\tau_{B}$ and for a fixed $\epsilon=0.001$. Notice that $n$ depend of $\tau_{B}$ and $\epsilon$, but has a direct dependence with the period $P$ and its derivative $\dot{P}$ too. In this case, when $t\rightarrow0,\:n\rightarrow 3-P\dot{P}^{-1}/\epsilon\tau_{B}$. The plot shows the behavior of the braking index for standard values of period and derivative of period of pulsars. Again, after a diffusion timescale $\tau_{B}$ it evolve like the familiar canonical model. That is, the growth of the magnetic field allow the braking index be less than the canonical value. After a timescale $t>\tau_{B}$, the braking index parameter tends to $n=3$.

\section{Conclusions}

We have revisited the magnetic torque problem from two different approaches: changes due growth of the magnetic field in the pulsar and changes in the moment of inertia.
In the first approach, we studied the early evolution of the magnetic field through a growth function $f(t)$, which has a exponential behavior, and followed its consequences on the overall properties of  RPP, focusing on very young pulsars.
The main motivation for the present study is due to the recent observations of a population of young pulsars, showing no magnetic field which is characteristic of these objects (\cite{Ho2011},  \cite{Magalhaes2012}). 
It has already been shown in previous work that the lack of magnetic field in young neutron star may be due to the fact that it is submerged into stellar crust during an early hypercritical phase. 

However, it is possible that when the hypercritical phase ends, this magnetic field may emerge to the surface, due to MHD instabilities or buoyancy effects, and be displayed as a delayed pulsar.
We find that the growth of the magnetic field change the early dynamics of the pulsar and modifies the RPP canonical model in the early regime. In particular, we observe that initially the characteristic age is higher than expected, depending on how the magnetic field grows. 
This implies that the pulsar looks older than it actually is. As was suggested by \cite {Geppert1999}, one way to identify such a pulsars would be through an association with a supernova remnant, which is a challenging task. 
Also, the initial spin-down luminosity is lower than that expected from the canonical model, however, it eventually increases until reaching the canonical behavior (again due to the growth of the magnetic field). This is interesting because if the surface magnetic field is low (in the early history of the pulsar), the system radiates its energy inefficiently, allowing a low initial spin-down luminosity and a high initial characteristic age. In contrast, the spin period evolves in similar way to the canonical model, due to its weak dependence on $f(t)$. Thus it remains practically unchanged for $t<\tau_{B}$ and a slight change is observed during the growth of $B$.

Finally, we show that the braking index is less than the expected value from the theoretical canonical model, $n=3$, when a growth of magnetic field process is present. We believe that such behavior may help explain (even if only partially) the low values observed in the few young pulsar where this parameter has been measured accurately. Nevertheless, very accurate values ​​of the age of the supernova remnant are necessary for a better estimate of diffusion timescale $\tau_{B}$. Of course, we are assuming that the RPP model is correct in general, and we augment it by including the possibility of magnetic field growth. 
%growth of the magnetic field in the early phase of the newborn neutron star to account the low values of the braking index through this mechanism.
 For instance, the Crab pulsar has, until now, the better initial spin determination, which was obtained independently: $P_{0}=19$ ms. The values of period and its derivative ($P=33$ ms and $\dot P_{-14}=42$ s s$^{-1}$) obtained from timing observation allow to estimate the current braking index as $n=2.51$ which tends very slowly to its canonical value. The current age is known from historical records (SN1054) and it is $t\simeq960$ yr. With these parameters it is possible to estimate the rotational energy as, $E_{rot}\simeq2\times10^{46}I_{45}P^{-2}(\mathrm{s})\: \mathrm{erg} \simeq2\times 10^{49}$ erg. The spin-down luminosity is given by, $\dot E_{rot}\simeq-4\times10^{32}I_{45}P^{-3}(\mathrm{s})\dot{P}_{-14}\: \mathrm{erg\: \mathrm{s}^{-1}}\simeq-5\times 10^{38}$ erg s$^{-1}$. The inferred superficial magnetic field is then, $B_{12}\simeq6 (I_{45}P(\mathrm{s})\dot{P}_{-14})^{0.5}/{R_{6}^{3}\sin\alpha}\: \mathrm{G} \simeq7$, where $I_{45}$ is the moment of inertia in units of $10^{45}\: \mathrm{g\: cm^{2}}$ and $\dot{P}_{-14}$ is the period derivative in units of  $10^{-14}\: \mathrm{s\: s{}^{-1}}$.
Also, if the magnetic field grew three orders of magnitude in such system ($\epsilon=0.001$), the estimated diffusion timescale is given by, $\tau_{B}=520$ yr. 
%Notice that the above values of period an its derivative were obtained from several sources, independently (\cite{Manchester2006}). But, in particular, was necessary to know the current age of the pulsar which is very difficult to obtain accurate for other pulsars.
\cite {Muslimov-Page1995} and \cite {Vigano-Pons2012} gave estimates of the difussion timescale performing 1D/2D numerical simulations.
Meanwhile, we do a phenomenological study of the behavior of the magnetic field and studies other properties of the pulsar.  We propose that it would be possible to estimate this parameter more accurately using the observed data for each pulsar with such simulations. This will accurately calculate the expected values ​​of $n$ showing that the growth of the magnetic field in these pulsars is feasible. This approach is prudent and necessary to carry out but is beyond the scope of this work.

\end{document}